\documentstyle[prb,aps,psfig]{revtex}
\begin{document}
\newcommand {\be}{\begin{equation}}
\newcommand {\ee}{\end{equation}}
\newcommand {\bea}{\begin{eqnarray}}
\newcommand {\eea}{\end{eqnarray}}
\newcommand {\nn}{\nonumber}

\draft
%
%
%
%

\title{
Excitation Spectra of Structurally 
Dimerized and Spin-Peierls Chains in a Magnetic Field 
}

\author{Weiqiang Yu and Stephan Haas}
\address{Department of Physics and Astronomy, University of Southern
California, Los Angeles, CA 90089-0484}

\date{\today}
\maketitle

\begin{abstract}
The dynamical spin structure factor and the Raman response are calculated
for structurally
dimerized and spin-Peierls chains in a magnetic field,
 using exact diagonalization techniques. 
In both cases there is a spin liquid phase composed of interacting
singlet dimers at small fields $h < h_{c1}$,
an incommensurate
regime ($h_{c1} < h < h_{c2}$) in which 
the modulation of the triplet excitation
spectra adapts to the applied field, and a fully spin polarized phase
above an upper critical field $h_{c2}$.
For structurally dimerized chains, the spin gap closes in the incommensurate
phase, whereas spin-Peierls chains remain gapped.
In the spin liquid regimes,
the dominant feature of the triplet spectra is a one-magnon bound state, 
separated from a continuum of states at higher energies.
There are also indications of a singlet bound state above the one-magnon
triplet.
\end{abstract}

\section{Introduction}

In the absence of a magnetic field,
structurally dimerized spin chains, such as 
$(VO)_2P_2O_7$\cite{garrett},
have a magnetic response similar to spin-Peierls chains, 
such as $CuGeO_3$\cite{arai,fujita}. 
In both cases, there is a one-magnon triplet bound state at 
energy transfers $\omega = \Delta_{z}$, where 
$\Delta_{z}$ measures the singlet-triplet spin gap,
followed by a two-magnon continuum of states with an onset at higher
frequencies, $\omega =       
2 \Delta_{z}$.\cite{uhrig}
The one-magnon bound state arises from the confinement of 
soliton-antisoliton pairs due to an effective attractive 
potential caused by the lattice dimerization.\cite{affleck,uhrig1}
In the case of spin-Peierls chains, there is additional coupling 
of the spins
to the elastic degrees of freedom of the lattice, causing the 
softening of a phonon mode.\cite{cross,lima,khomskii} 
In contrast, the atomic positions of structurally
dimerized chains are completely locked, ruling out a feedback 
between the spins and the phonons. It is this magneto-elastic feedback 
in the spin-Peierls compounds,
which allows the spin gap to remain open even at large magnetic 
fields,\cite{uhrig2}  while in the structurally
dimerized chains, the gap closes beyond a critical
magnetic field strength, $h_{c1}$, due to the deconfinement of the
soliton-antisoliton pairs. 
In both cases the triplet
spectra become incommensurate at $h > h_{c1}$.
However, we will see that for the structurally dimerized 
chains they are gapless soliton-antisoliton continua, 
whereas in the spin-Peierls
compounds the dominant one-magnon bound state remains gapped, 
acquiring a modulation which depends on the magnitude of the applied field. 
\cite{uhrig2}

In the adiabatic approximation (suppressing the phonon dynamics), the 
effective Hamiltonian of antiferromagnetically correlated spins
coupled to a crystal lattice is given by,
\bea
H = J \sum_{r=1}^N (1 + \delta_r) {\bf S}_r \cdot {\bf S}_{r+1} 
+ \frac{K}{2} \sum_{r=1}^N \delta_r^2,
\eea
where $J$ is the Heisenberg exchange constant, $\delta_r$ are local
lattice distortions, and $K$ is the lattice spring constant. The feedback
of the phonons to the spin degrees of freedom is contained in the dependence
of $\delta_r$ on $K$. In the absence of a magnetic field, $\delta_r
= \delta (-1)^r$. The dimerization parameter $\delta$ is a constant
for structurally dimerized chains, whereas $\delta \propto K^{-3/2}$ for
spin-Peierls systems.\cite{cross,spronken,feiguin}
The structurally dimerized chain can be viewed as a
limiting case of Eq. (1) with a vanishing spring constant
($K = 0$)
and a ``frozen" (inelastic) lattice modulation, which does not vary with
an applied magnetic field. On the contrary, spin-Peierls chains in
a sufficiently
high magnetic field, $h > h_{c1}$, gain elastic energy by adjusting their
lattice to the field.\cite{lima,uhrig2,bray,kiryukhin,schonfeld}
This magneto-elastic distortion can be rather well 
approximated by a sinusoidal form,
$\delta_r = \delta \cos{(q r)}$, with 
$q = \pi + 2\pi \langle S^z_{tot}/N \rangle$.\cite{uhrig2,schonfeld,footnote1} 
With these two modulations of $\delta_r$, the magnetic field induced
transition from the dimerized to the incommensurate phase is 
continuous for structurally dimerized chains, while it is first order
for spin-Peierls chains, where 
a jump of $K \delta^2/4$ occurs in the elastic energy.\cite{schonfeld} 

In this work, the spin excitation spectra in a magnetic field of these
two systems are contrasted. Using
exact diagonalization techniques on finite lattices of up to $N=24$ sites
with periodic boundary conditions, the
triplet and singlet excitations are calculated, allowing a direct comparison
with inelastic neutron and Raman experiments. 
The strength of
this method, although restricted to relatively small lattice sizes, is
the accessibility of the full excitation spectrum for a given cluster.
As our interest is primarily in the magnetic response, we concentrate on
two effective, purely magnetic model Hamiltonians, $H_{dim}$
and $H_{sP}$, derived from $H$ (Eq. 1). Phonon contributions, other 
than entering via the parametrization of $\delta_r$ 
will be neglected. 

For the structurally dimerized spin-1/2 antiferromagnetic Heisenberg chain, 
the model Hamiltonian $H$ reduces to
\bea
H_{dim} = J \sum_{r=1}^N (1 + \delta (- 1)^r) {\bf S}_r \cdot {\bf S}_{r+1}.
\eea
In a quasi-one-dimensional compound, such as $KCuCl_3$,
the dimerization parameter, $\delta > 0$, 
originates from the alternating spacing
of the spin-carrying $Cu^{2+}$ ions.\cite{cavadini}
In the limit $\delta = 1$, the system is an ensemble of $N/2$
uncoupled dimers with only two energy levels per dimer, and a spin gap
$\Delta_{z} = 2 J$. 
In the opposite 
limit $\delta = 0$, $H_{dim}$ reduces to the
isotropic antiferromagnetic Heisenberg chain which is
quasi-long-range ordered, and thus belongs to a different
universality class from the dimerized system.
For sufficiently small lattice dimerizations,
a regime with sizeable inter-dimer interactions can be identified with
the scaling properties of the massive Thirring model.\cite{affleck}

The effective magnetic Hamiltonian for
spin-Peierls compounds is given by\cite{uhrig2}
\bea
H_{sP} = J \sum_{r=1}^N (1 + \delta \cos{(q r)}) {\bf S}_r \cdot {\bf S}_{r+1}.
\eea
Here, the feedback due to the interactions of the spins with
the lattice phonons enters through a field dependent modulation of
the effective nearest-neighbor exchange integral, $J_{eff}(r) =
J ( 1 + \delta \cos{(q r)})$, where
$q = \pi + 2\pi \langle S^z_{tot}/N \rangle$. 
In the commensurate phase ($h < h_{c1}$), the modulation is fixed
at $q = \pi$, and $H_{sP}$ is identical to $H_{dim}$.
However, in the incommensurate regime 
$q$ continuously grows from $\pi$ to $2\pi$, 
mimicking the elastic distortion of the underlying lattice due to
the dynamical coupling with the spins. Hence, all eigenstates of this system 
have a spin as well as a phonon component.\cite{uhrig0} 

In the subsequent section, the spin excitation spectra of structurally
dimerized chains ($H_{dim}$)
in a magnetic field are discussed, followed by a section on spin-Peierls
systems ($H_{sP} $). 
We finish with some concluding remarks.
As we are interested
in capturing the generic features, and in particular the differences, of the
two physical situations described above, no compound specific
parameters, such as  inter-chain or next nearest-neighbor
exchange couplings, are considered.
Rather, our focus will be on understanding the characteristic features of
the phases in the most elementary magnetic models.

\section{Structurally Dimerized Chain}

It is remarkable that for some quasi-one-dimensional compounds,
such as $(VO)_2P_2O_7$ and $CuGeO_3$, it has been quite difficult to establish
a unique microscopic model.\cite{garrett,johnston,castilla,barnes}
For example, from fits to 
early measurements of the uniform susceptibility 
on $(VO)_2P_2O_7$ it has been concluded that this material 
is either a structurally dimerized or a frustrated spin-1/2 Heisenberg 
chain with a sizeable next-nearest-neighbor exchange coupling. 
\cite{johnston,barnes}
In both cases, a spin gap opens up either due to a structural or to a frustration
induced dimerization, and the resulting
thermodynamic response is quite similar for the two proposed models.
Only recently, it could be shown by 
inelastic neutron scattering spectroscopy that $(VO)_2P_2O_7$ is indeed a 
structurally dimerized chain, and that it is the lattice distortion rather
than any frustration which gives rise to the observed spin gap.
\cite{garrett}
Similarly, there
are still rather different effective
parameter sets for $\delta$ and $J_2$ used in the 
current literature on the spin-Peierls compound $CuGeO_3$. In one case
$(\delta= 0.03, J_2 = 0.24)$ 
the spin gap opens because of the lattice distortion
\cite{castilla},
while in the other case $(\delta = 0.014, J_2 = 0.36 )$ 
the frustration alone is large 
enough to cause a spin gap.\cite{riera1}  
It is thus of particular interest to examine the full spin excitation spectrum
of these quasi-one-dimensional materials in a magnetic field, in order to 
pinpoint the most relevant microscopic interactions.

Let us start by discussing 
the phase diagram of the structurally dimerized spin-1/2 Heisenberg chain in
a magnetic field, shown in Fig. 1. 
(i) for $|h|<h_{c1}$, the system is in a
spin-liquid phase with a singlet-triplet
spin gap $\Delta_{z}$
and  $h_{c1}= \Delta_{z} $;
(ii) for $h_{c1}<|h|<h_{c2}$ it is a gapless spin-density wave with
a field-dependent modulation;
and (iii) for $|h|>h_{c2} = 2 J$ it is fully spin-polarized
in the direction of the applied magnetic field.
To determine the dependence of $\Delta_{z}$ on the dimerization,
we use  Shanks' transformation\cite{shanks,sakai} 
on lattices of up to N=24
sites. The asymptotic form of the spin gap for a given dimerization 
obeys a finite-size scaling relation,
\bea
\Delta_{z}(N,\delta) = \Delta_{z}(N=\infty,\delta ) + A(\delta) 
\exp{(-\Gamma(\delta) N)},
\eea
where the constants $A(\delta)$ and $\Gamma(\delta)$ are obtained   
from Shanks' recursive   
equations\cite{shanks,sakai}.
In accordance with Ref. \cite{barnes1}, we find that the initially proposed  
dependence, $\Delta_{z}(N=\infty ,\delta ) \propto
\delta^{2/3} /\sqrt{|\log{\delta}|},$\cite{cross,black,klumper} matches our 
extrapolation rather poorly, while the form 
$\Delta_{z}(N=\infty ,\delta ) = 2 \delta^{3/4}$ (shown in Fig. 1)
gives an excellent fit
to our data over the whole range of parameter space, $\delta \in (0,1]$. 
One likely reason for this discrepancy  is that the initial analytical
prediction is valid only for very small values of $\delta$, 
difficult to access with a finite-size scaling procedure.
Furthermore, in this regime higher
order logarithmic corrections also become important.\cite{affleck2}
Down to rather small values of $\delta$ ($\delta > 0.3$), the dependence
of the spin gap is linear to leading order,
$\Delta_{z} \approx (1 + 3\delta)/2$, indicating
that the picture of weakly interacting dimers - a perturbation
about the limit of isolated dimers ($\delta = 1$) -  is applicable in this 
parameter regime. However, at low values of the 
dimerization parameter ($\delta \alt 0.3$) the interactions between the
dimers become increasingly important, leading to a deviation from the
linear dependence of the spin gap on $\delta$.

\begin{figure}
\centerline{\psfig{figure=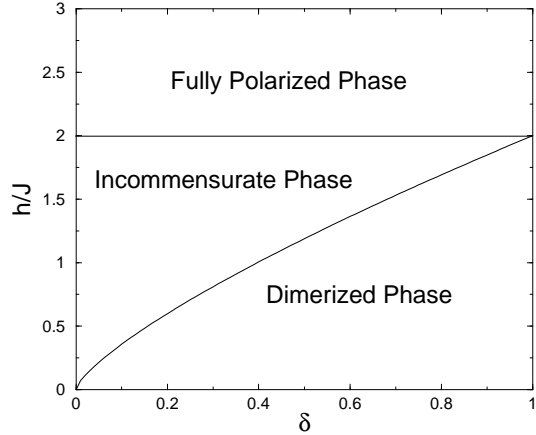,width=7cm,height=6.0cm,angle=0}}
\caption{
Phase diagram of the structurally
dimerized spin-1/2 Heisenberg chain in a magnetic
field. $\delta$ is the dimerization parameter.
}
\end{figure}

The energy 
gaps between the groundstate and the lowest excited singlet
and triplet as a function of the lattice dimerization are shown in Fig. 2. 
The magnitudes of the 
gaps in this figure quantitatively resemble the thermodynamic 
limit ($N \rightarrow \infty$), as they were obtained from
Shanks' transformation.\cite{footnote2} In the limit of vanishing 
dimerization, the gaps disappear, indicating that the groundstate is
qualitatively different for the cases of vanishing and finite dimerization.
The singlet gap is always larger than the triplet gap, and their ratio
is $\Delta_S/\Delta_{z} = 2$ for most of parameter space.
\cite{gros,bouzerar} However,
at lower values of $\delta$ this ratio becomes smaller, possibly 
approaching the predicted value of $\sqrt{3}$ as $\delta \rightarrow 0$.
\cite{uhrig,augier}
Unfortunately, the quality of
our finite size extrapolation procedure deteriorates in this limit,
and no definite confirmation of $\Delta_S/\Delta_{z} = \sqrt{3}$
can be drawn from this study, although our data is consistent with this value.
It is clear, however, that the region of small dimerization must 
be governed by
a field theory such as the massive Thirring model with a non-linear 
dependence of the excitation gaps on the effective mass,
which in turn is proportional
to $\delta$.\cite{affleck}

\begin{figure}
\centerline{\psfig{figure=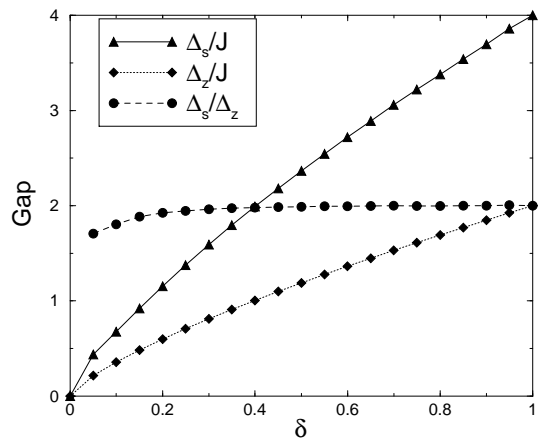,width=7cm,height=6.0cm,angle=0}}
\vspace{0.5cm}
\caption{
Singlet and triplet spin gaps in the structurally
dimerized spin-1/2 Heisenberg chain 
at zero magnetic field, obtained by Shanks' transformation on lattices
of up to N=24 sites. (i) diamonds: singlet-triplet gap, (ii)
triangles: singlet-singlet gap, (iii) circles: ratio of singlet-singlet
and singlet-triplet gap.
}
\end{figure}

In Fig. 3, the triplet excitation spectra of the structurally
dimerized spin-1/2
Heisenberg chain are shown, along with the corresponding dispersion 
relations
in the insets. These spectra were calculated by an exact numerical
diagonalization of 
18-site chains with periodic boundary conditions, combined with a
continued fraction expansion to obtain the full dynamical response
functions. Let us first examine the dynamical spin structure factor,
\bea
S^{zz}(k,\omega) = \sum_n |\langle n
 | S^z_k | 0 \rangle|^2 
\delta(\omega - E_n + E_0),
\eea 
where $S^z_k = \frac{1}{\sqrt{N}} \sum_r \exp(i k r) S^z_r$ is the 
projection of the spin operator
parallel to the applied magnetic field, $ | n \rangle$
denotes an eigenstate of the Hamiltonian with energy $E_n$, orthogonal
to the groundstate with energy $E_0$,
and magnetization $m = \langle S^z_{tot}/N \rangle
\in [0,1/2]$.
In the dimerized phase (Fig. 3(a)), the dominating feature in the spectrum is
the one-magnon bound state
with an onset frequency $\omega = \Delta_{z} \approx (1 + 3\delta )/2 $,
well separated from a continuum of states starting at twice
this energy. Increasing the magnetic field, the one-magnon bound state moves 
down to lower energies, and eventually the gap closes at $h = h_{c1}$
(Fig. 3(b)). Beyond $h_{c1}$, the soliton-antisoliton confinement
potential is thus
overcome by the magnetic field, and the bound state decays into 
a low-energy two-spinon continuum, similar to the spin-1/2 Heisenberg chain.  
In addition, there are continua of states at higher
energies. In particular the lowest continuum (starting at $\omega 
= \Delta_{z}$)
carries most of the spectral weight.
At the onset of the incommensurate phase ($h = h_{c1}$, Fig. 3(b)) 
the phase space of this low-energy continuum is strongly restricted,
and therefore its width is small. The reason
will become clear from the discussion below in terms of the 
corresponding spinless fermion picture.
As the applied field is increased from $h_{c1}$ to $h_{c2}$,
the width of the low-energy continuum grows, and
the wave vector of the dominant infrared divergence
moves continuously from $q = \pi$ to 
$q = 2 \pi$. In Fig. 3(c), the triplet excitation
spectra are shown at a particular magnetization,
$m = 4/18$, corresponding to a magnetic field
$h\simeq 1.58 J$. Clearly, the modulation at this
field is incommensurate, and the magnetic unit cell is enlarged by
approximately a factor of two with respect to its size at zero field.
Furthermore, the phase space for triplet 
excitations is reduced with increasing magnetic field,
leading to an overall loss of spectral weight at higher fields.
Finally, close to $h_{c2}$ two triplet bands emerge,
split by a dimerization gap, $\Delta_{\pm} = 2\delta$ (Fig. 3(d)). 
Low-energy spectral weight
away from long wavelengths disappears, and the low-frequency
dispersion approaches $\omega  \propto k^2$, characteristic for 
ferromagnetism.

\begin{figure}
\centerline{\psfig{figure=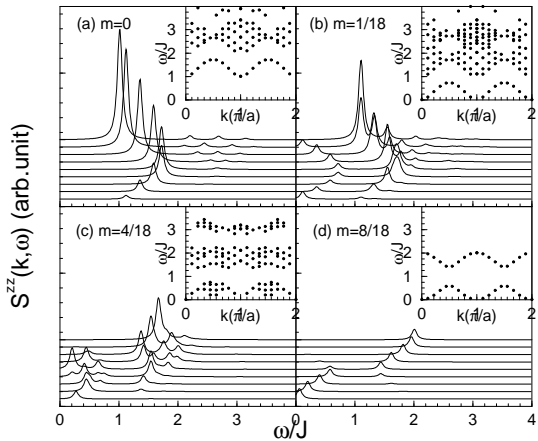,width=7cm,height=6.0cm,angle=0}}
\vspace{0.5cm}
\caption{
Triplet excitation spectra of the structurally
dimerized spin-1/2 Heisenberg chain in a 
magnetic field, calculated on an 18-site lattice with $\delta = 0.4$. 
The lowest curve is the dynamical spin structure factor
at momentum transfer $k = 0$,
and the highest at $k = \pi$.
(a) Dimerized regime ($h =0$), 
(b) onset of incommensurate phase ($h = h_{c1}$), (c) center of 
incommensurate phase ($h_{c1} < h < h_{c2}$) ,
(d) incommensurate phase at a high magnetic
field, close to 
$h_{c2}$. 
The corresponding pole positions are shown in the insets.
}
\end{figure}

The Hamiltonian of the 
structurally dimerized spin-1/2 Heisenberg chain can be mapped
onto a model of spinless fermions via the Jordan-Wigner transformation.
Due to the lattice dimerization, the spinless fermion band is split
into two parts which disperse according to
\bea
\omega_{\pm}(k)/J = 1 \pm \sqrt{\delta^2  + (1 - \delta^2) \cos^2(k) },
\eea
giving rise to a dimerization
gap, $\Delta_{\pm} = 2 \delta$, and to a total single particle
bandwidth of $2J$. 
While these dispersion relations are
exact in the XY-limit of $H_{dim}$, they are 
only slightly renormalized 
in the isotropic Heisenberg limit for sufficiently large dimerization values
($\delta > 0.3$), as can be seen by comparing $\omega_-(k)$ and 
$\omega_+(k)$
with the dispersions in the inset of Fig. 3(d).
Furthermore, in the spinless fermion picture, 
the applied field corresponds to a
chemical potential. At vanishing magnetic field, the chemical potential 
lies in the center of the gap between $\omega_-$ and $\omega_+$. 
In order to excite an unbound
pair of particles, a minimum energy of $2 \Delta_{\pm}$
is needed. In addition, due to the attractive scattering between
the spinless fermions, an exciton-type 
particle-hole bound state is formed with a minimum
energy of $\Delta_{z}(h)$, dispersing at $h = 0$ as
\bea 
\omega_z(k)/J
=  (1 + \delta) - (1 - \delta) \cos (2 k)/2 ,
\eea
as observed in the inset of Fig. 3(a).\cite{uhrig} At small magnetic fields
($h < h_{c1}$) this dominant one-magnon triplet mode
carries most of the weight in the dynamical structure factor. 
Furthermore, there is a 
second gap ($\Delta_2 = \Delta_z$) between the one-magnon bound state and 
a continuum of states which is a simple convolution of two magnons
with a dispersion $\omega_z(k)$.
With increasing magnetic field ($h \rightarrow h_{c1}$) the bound state moves
down to lower energies, and eventually $\Delta_z$ vanishes at $h_{c1}$,
whereas the onset of the continuum now occurs at $\Delta_2(h=h_{c1})
= \Delta_z(h=0)$. Beyond $h_{c1}$ the one-magnon bound state disappears
and decays into a particle-hole continuum as the effective confining potential
is overcome by the applied field.

Using only  $\omega_z(k)$ and $\omega_{\pm}(k)$, the complete
magnetic field dependence of the triplet spectra can thus be understood 
qualitatively within a simple rigid-band picture.
In the incommensurate phase, the chemical potential moves into the
lower band, $\omega_-(k)$. The continuum of states at low energies
arises from two-particle excitations within $\omega_-(k)$,
whereas the continua at higher frequencies stem from processes involving
interband scattering.
The modulation wave vector $q$,
corresponding to the applied magnetic
field $h$, is obtained from the solution of $h = \omega_-(q)$. 
At fields slightly above $h_{c1}$, the phase space for intraband 
scattering processes are restricted
within the lower band which is almost full. This is the reason for the 
narrow width of the low-energy continuum in Fig. 3(b), just at $h=h_{c1}$. 
At large magnetic fields, $h \alt h_{c2}$, the phase space of triplet 
excitations is exhausted, and the dynamical structure factor traces out 
the single particle bands of the spinless fermions (Fig. 3(d)).

Because the spectra in Fig. 3 were obtained on finite-size lattices, there are
no true branch cuts in $S^{zz}(k,\omega)$. Rather, discrete sets of poles 
appear where continua are expected to emerge in the thermodynamic limit.  
In order to distinguish bound state poles from sets of poles which 
become part of a continuum in the thermodynamic limit,
a finite-size scaling analysis of the 
individual pole positions and weights is necessary. From such an 
extrapolation, using chains with N = 4, ... , 24 sites, we find that the
bound state with $\omega_z(k)$ (Fig. 3(a)) is indeed stable, whereas
the other poles in the spectrum merge
into continua as the lattice size is increased to infinity. 
At smaller values of the lattice dimerization ($\delta \alt 0.3$), the
general features of the triplet spectra are the same as discussed above.
However, once the bandwidths of $\omega_{\pm}(k)$ become larger than 
the dimerization gap separating the two bands, $\Delta_{\pm}$, 
the higher-energy continua, which are
separated from each other for larger dimerizations, begin to overlap. Using the
exact dispersion expressions for the XY limit (Eq. 6), this crossover 
occurs at $\delta_c = 1/3 \approx 0.3$, 
consistent with the deviations from the weakly
interacting dimer picture observed in our numerical data.

\begin{figure}
\centerline{\psfig{figure=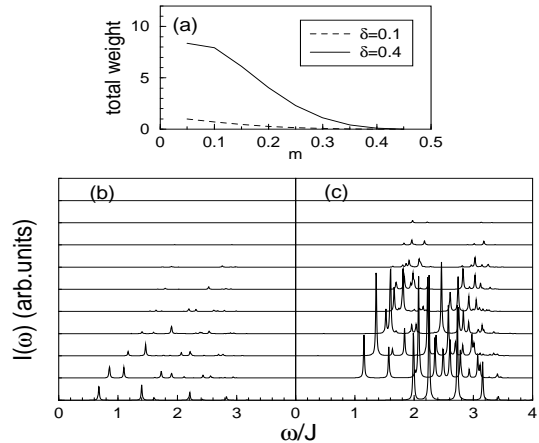,width=7cm,height=6.0cm,angle=0}}
\vspace{0.5cm}
\caption{
Raman spectra for a 20-site
spin-1/2 structurally dimerized Heisenberg chain in a magnetic
field with (b) $\delta = 0.1$ and (c) $\delta = 0.4$. The integrated
weight for these two cases is shown in (a). The spectra are for magnetizations
$m=0$ (lowest curve) up to $m = 9/20$ (top curve). 
}
\end{figure}

Let us now turn to the spin excitation spectra of the structurally
dimerized spin chain in a magnetic field, as they are probed 
by Raman scattering measurements. Within the Loudon-Fleury 
theory\cite{loudon} - 
assuming resonant scattering - the effective $A_{1g}$ Raman operator of a
one-dimensional spin system is proportional to 
$\sum_r {\bf S}_r \cdot {\bf S}_{r+1}$. Setting the proportionality
constant equal to one, the dynamical Raman response function takes the
form 
\bea
I(\omega ) = 
 \sum_n |\langle n
 | \sum_r {\bf S}_r \cdot {\bf S}_{r+1} | 0 \rangle|^2 
\delta(\omega - E_n + E_0).
\eea
In the following discussion of $I(\omega )$,
terms in the Raman operator which are proportional to the
Hamiltonian are omitted.\cite{footnote3}
In Fig. 4 the Raman spectra for 20-site chains are shown as a 
function of the magnetic field. At zero magnetization ($h < h_{c1}$),
one singlet bound state is expected with a gap $\Delta_s$, followed by a
continuum of excitations at higher energies, involving 4 spinons.\cite{augier}
In the regime of large dimerization ($\delta > \delta_c 
$), the singlet
gap is at $\Delta_s = 2 \Delta_z \simeq (1 + 3\delta)J$ (Fig. 4(c)). 
This
bound state is best understood by considering the
limit of complete dimerization ($\delta = 1$). The corresponding groundstate
at $h = 0$ is a product of singlet dimers with energy $-3 J/2$ per
dimer.\cite{barnes1}
Two such dimer singlets can be excited by the Raman operator
into a 4-site singlet state.
The energy difference, $\Delta_s$, between these states is approximately
equal to $(1 + 3\delta)J$ as long as the dimer-dimer interactions are 
sufficiently small ($\delta > \delta_c$). In the limit 
of complete dimerization, one obtains exactly $\Delta_s = 4J $. 
In the opposite limit,
the singlet bound state moves to lower energies (Fig. 4(b)), and most of the
spectral weight is transferred into the zero-frequency peak, 
not shown here. Apart from the bound state at $\Delta_s$, there is a 
continuum of excitations at higher energies. Evidently, the spectra 
in Fig. 4 are plagued by severe finite-size effects, such that it is
particularly difficult to distinguish by inspection the 
precursors of continua from emerging isolated bound states.
\cite{footnote4}
However, from the 
scaling behavior of the individual pole positions and 
weights we conclude that our data at zero magnetic field is consistent with
the picture of an isolated bound state at $\Delta_s$,
followed by a continuum of states
above a 
threshold $\omega \agt \Delta_z$ in the thermodynamic limit. 
Beyond $h_{c1}$, the bound state disappears, and the onset frequency of the 
continuum increases from $\Delta_z$ at $h_{c1}$ up to $2J$ close
to $h_{c2}$. In Fig. 4(a) the integrated weight of the Raman spectrum,
$W = \int d\omega I(\omega)$,  
is plotted as a function of the magnetization. $W$ becomes
very small as $\delta \rightarrow 0$. Furthermore, with increasing
magnetic field, $W(m)$ decreases more rapidly - with a purely concave
shape -  in the case of small dimerization (Fig. 4(b)). 
However, this subtle difference
is most likely of little experimental relevance because important
compound specific
contributions, such as frustrating longer range interactions or
interchain coupling, have been
neglected in this discussion.

\section{Spin-Peierls Chain}

\begin{figure}
\centerline{\psfig{figure=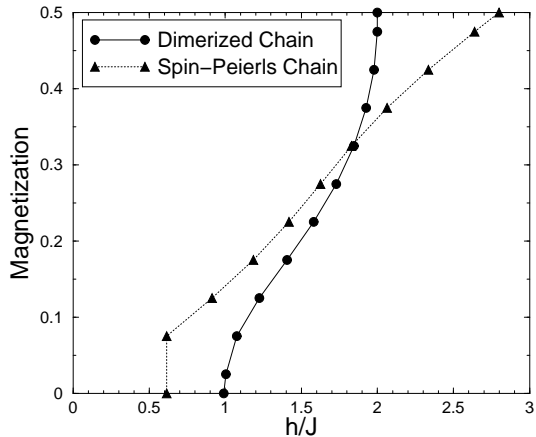,width=7cm,height=6.0cm,angle=0}}
\vspace{0.5cm}
\caption{ Magnetization curves of the spin-1/2 structurally
dimerized Heisenberg
chain (circles) and of the spin-Peierls chain (triangles)
with $\delta = 0.4$. At $h_{c1}$, the magnetization of the spin-Peierls
system jumps discontinuously from zero
to a finite value, indicating a first order
transition, whereas in the structurally 
dimerized chain the transition into the 
incommensurate regime is 
continuous.
}
\end{figure}

In contrast to the structurally dimerized chains, spin-Peierls chains have 
a gapped incommensurate phase as their lattice
distortion adapts
magneto-elastically to the applied field.
For a self-consistent treatment of this
incommensurate regime, the phonon and the spin degrees of freedom 
have thus to be treated
on an equal footing. In large scale numerical studies of the full
adiabatic spin-phonon Hamiltonian $H$ (Eq. 1), it has been 
shown that the structural distortion of the lattice and the modulation 
of the local magnetization have the shape of solitons, natural for 
one-dimensional systems.\cite{schonfeld,feiguin} As the focus of this work is
the magnetic response, 
an effective parametrization for the lattice distortion
of the form $\delta_r = \delta \cos(qr)$
is used,\cite{uhrig2}
instead of treating the elastic part of $H$ self-consistently.
Especially at higher magnetic fields,
this form gives results almost identical to the
self-consistent treatment of the lattice dynamics 
for observables such as the local magnetization.
Furthermore, we have
verified that other (solitonic) parametrizations\cite{footnote1}
yield spin excitation spectra which are only 
minutely different to those with a sinusoidal lattice distortion.

In Fig. 5, the magnetization curves, $m(h)$,
of the spin-1/2 structurally dimerized Heisenberg
chain and of the spin-Peierls system are shown. They were calculated
by numerical diagonalization of 
22-site chains. The magnetization of
the structurally dimerized chain has a plateau between $h = 0$
and $h_{c1} = \Delta_{z}$. $m(h)$ then
rises continuously from zero at $h_{c1}$ up to $m=1/2$ at $h_{c2} 
= 2J$. The particle-hole symmetry of the corresponding spinless fermion
band is reflected by the point symmetry of $m(h)$ about its midpoint.
In contrast, the magnetization of the spin-Peierls chain 
jumps discontinuously at $h_{c1}$ from zero to a finite value. The 
position of $h_{c1}$ and the magnitude of the discontinuity depend on the
lattice spring constant $K$. Therefore, a precise
determination of these quantities is beyond the realm of our effective
- purely magnetic - theory using $H_{sP}$. However, from self-consistent
calculations it has been concluded that
the second order transition line, $h_{c1}(\delta )$,
for the structurally dimerized chain (Fig. 1)
becomes first order in the spin-Peierls case, and moves down towards lower
fields with increasing lattice spring constant.
\cite{lima,schonfeld,bursill}

\begin{figure}
\centerline{\psfig{figure=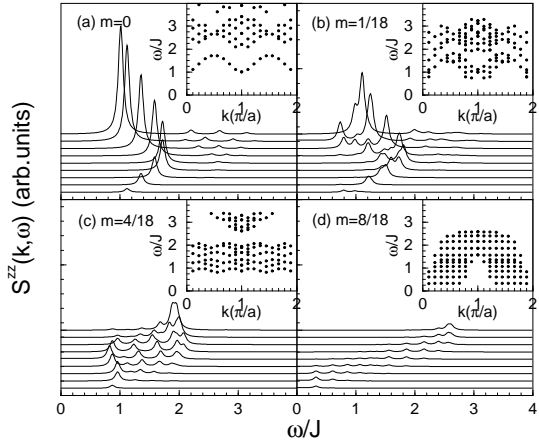,width=7cm,height=6.0cm,angle=0}}
\vspace{0.5cm}
\caption{
Triplet excitation spectra of the spin-1/2 spin-Peierls chain in a 
magnetic field, calculated on an 18-site lattice with $\delta = 0.4$. 
The lowest curve is the dynamical spin structure factor
at momentum transfer $k = 0$,
and the highest at $k = \pi$.
(a) Dimerized regime ($h =0$), 
(b) onset of incommensurate phase ($h = h_{c1}$), (c) center of 
incommensurate phase ($h_{c1} < h < h_{c2}$) , 
(d) incommensurate phase at a high magnetic
field, close to 
$h_{c2}$. 
The corresponding pole positions are shown in the insets.
}
\end{figure}

Let us now turn to the triplet excitation spectrum of the spin-Peierls 
chain in a magnetic field, shown in Fig. 6. 
For $h < h_{c1}$, the spin response is 
identical to that of the structurally
dimerized chain (Fig. 6(a)). The lattice dimerization gives rise
to a scattering potential peaked at $q = 2 k_F
= \pi$. This attraction between pairs of
particles at opposite ends of the Brillouin
zone leads to the formation of a one-magnon bound state.
In the incommensurate phase (Fig. 6(b-d)), the lattice
distortion,
$q = \pi + 2 \pi m(h)$, adapts to the magnetic field, thus 
supporting a bound state with a minimum energy 
$\omega = \Delta_{z}(h) \ne \Delta_{z}(h=0)$ and
a modulation $q$ due to an effective field dependent potential, 
$V_{eff} (q) \sim   \delta \cos(q r)$. At higher energies, there 
are continua of states. In contrast to the dimerized
phase, in the incommensurate regime the second gap, between the onset of the 
bound state and the onset of 
the lowest continuum, is smaller than $\Delta_{z} (h)$. This can be 
understood within the spinless fermion picture: as the magnetic field
is increased beyond $h_{c1}$, the corresponding Fermi wave vector moves
away from $\pi/2$. Opposite to the case of the structurally 
dimerized chain, the scattering potential adapts to the changing 
magnetic field, such that there is always an instability at the Fermi
level due to particle-hole
scattering with momentum transfer $q = 2 k_F(h) = \pi + 2 \pi m(h)$.
Furthermore, in the incommensurate phase 
the chemical potential is offset from the
center of the gap. The energy difference between $\mu$ and the lower
edge of the gap is $\Delta_-(h)$, and between $\mu$ and the upper edge
it is $\Delta_+(h)$, shown in Fig. 7, where the
values of $\Delta_-$ and $\Delta_+$
have been evaluated by acting with $S^-$ and $S^+$ on 
the groundstate.\cite{uhrig2} 
The onset of the one-magnon bound state occurs at 
a higher energy, $\Delta_{z}(h) >  (\Delta_-(h) + \Delta_+(h))/2$. 
Only in the dimerized phase, it is found that
$\Delta_{z} = \Delta_- = \Delta_+$, and thus 
$\Delta_{z} =  (\Delta_- + \Delta_+)/2$ because of the additional 
particle-hole symmetry for the special case of half-filling. 
The magnetic field dependence of the triplet spectra in Fig. 6 follows
exactly this picture. For example, at magnetization $m = 4/18$ (Fig. 6(c)),
the onset of the one-magnon bound state is at $\omega = 0.82 J$, whereas
the lowest continuum of states starts at $\Delta_- + \Delta_+ = 
0.39 J + 0.69 J = 1.08 J$, thus separating the bound state from the 
continuum by a second gap of $\Delta_2 = 0.26 J < \Delta_{z}(h) = 0.82 J$. 

Similar to the structurally dimerized chain, the phase space of 
triplet excitations is gradually reduced in the incommensurate phase
of the spin-Peierls system, 
leading to a loss of spectral weight in $S^{zz}(k,\omega)$  
with increasing magnetic field. Furthermore,
the widths of the triplet bands shrink at higher
fields, indicating an effective localization of triplets. For example, 
the band width $W$ of the 
one-magnon bound state is drastically reduced (see inset of Fig. 7), going
practically to zero beyond $m_c \approx 0.3$. 
Its dispersion, $\omega_z(k)$, oscillates
rapidly as the modulation vector $q$ approaches $2 \pi$.
The reason for this behavior is that the real-space magnetic unit
cell grows with increasing magnetic field. At sufficiently high
fields, it grows beyond the size of any finite cluster. For the parameter
choice and lattice size we use, this happens at approximately
$m_c$. Beyond $m_c$, the corresponding modulation of the 
effective nearest-neighbor exchange integral, $J_{eff}(r) =
J ( 1 + \delta \cos{(q r)})$, has only one minimum in the finite
chain where triplets are
trapped, leading to a ``smearing" of $\omega (k)$ in momentum space
(Fig. 6(d)). While this localization of magnons
is obviously an artefact of the finite cluster
calculation, it may be realized in mesoscopic chains, as soon as the 
size of the magnetic unit cell exceeds the mesoscopic length scale.
Also, such a localization can 
easily be stabilized by a pinning of the distortion,
to which a physical system is highly susceptible as the modulation 
grows toward infinity. In this case, a real-space picture of (almost) 
localized triplets is most appropriate at high magnetic fields, 
close to $h_{c2}$.
  
\begin{figure}
\centerline{\psfig{figure=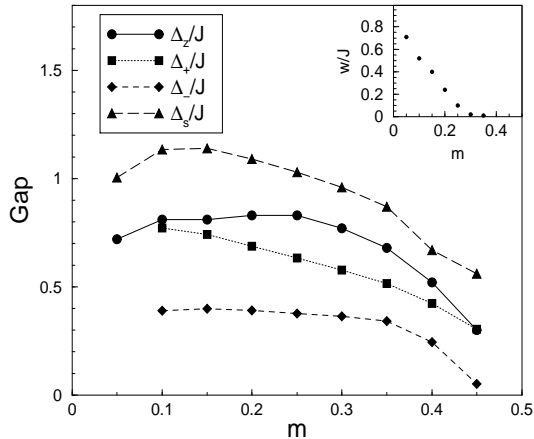,width=7cm,height=6.0cm,angle=0}}
\vspace{0.5cm}
\caption{
Singlet and triplet gaps in the incommensurate phase of the spin-Peierls
chain with $\delta = 0.4$, calculated on a 20-site lattice. (i) circles:
triplet gap parallel to the applied field, (ii) squares and diamonds:
triplet gaps perpendicular to the applied field, (iii) triangles: 
singlet gap.
}
\end{figure}

The singlet gap $\Delta_s$, shown in Fig. 7,
corresponds to the onset of the finite
frequency Raman spectrum (Fig. 8). It is always
larger than the 
triplet gap, and has a  
similar dependence on the magnetization.
From an examination of the finite-size scaling behavior of the
poles in the Raman spectrum, there appears to be a low-energy singlet
bound state for all fields $h < h_{c2}$.\cite{uhrig,augier} 
In particular, an analysis
of the $\delta =0.4$ Raman spectra
(Fig. 8(c)) suggests that in the thermodynamic limit the two lowest poles
merge into one, and their spectral weight extrapolates to a finite 
value, thus indicating the existence of a two-magnon bound state. For smaller
dimerizations (such as $\delta = 0.1$ in Fig. 8(b)) it is difficult
to determine from our finite-size
data whether there is a bound state. This is
consistent with a recent numerical study of the Raman spectrum 
in $CuGeO_3$ which has considered
an even smaller dimerization constant $\delta \simeq
0.03$.\cite{gros} Here it was argued that the strong magnon-magnon interactions 
destabilize the singlet bound state. However, as
seen in Figs. 8(b) and (c), the Raman excitation spectra are qualitatively
rather similar for these two parameter choices. 
Consider for example the spectra at the onset of the incommensurate
phase ($m = 1/20$), shown in the second lowest curves of 
Figs. 8(b) and (c). At low energies, there is a pair of poles (the second
pole is not visible for $\delta = 0.1$), separated
from a set of poles at higher energies with a close spacing.
As discussed above, the low-frequency poles merge in the
thermodynamic limit, whereas the high-frequency set of poles appears to 
evolve into a continuum of states.
We therefore suspect that singlet bound states may exist at the lower
edge of the spectrum for any finite $\delta$, but much larger clusters
may be needed for a numerical confirmation of the singlet bound state at
small values of $\delta$. Furthermore, the total finite-frequency
spectral weight $W(m)$, shown in Fig. 8(a), has the same dependence
on the magnetization for both choices of $\delta$, indicating that there 
should be only one massive field theory applicable for the whole range
$\delta \in (0,1]$. 

\begin{figure}
\centerline{\psfig{figure=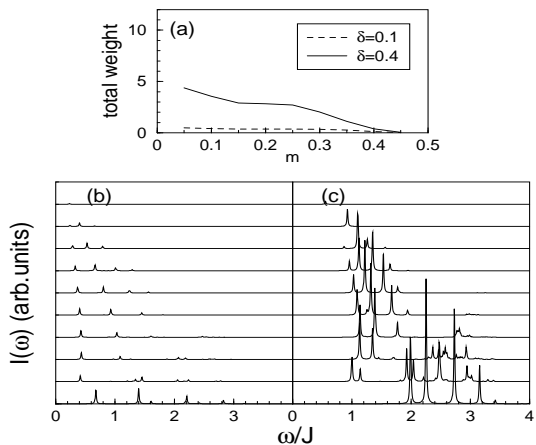,width=7cm,height=6.0cm,angle=0}}
\vspace{0.5cm}
\caption{
Raman spectra for a 20-site
spin-Peierls chain in a magnetic
field with (b) $\delta = 0.1$ and (c) $\delta = 0.4$. The integrated
weight for these two cases is shown in (a). The spectra are for magnetizations
$m=0$ (lowest curve) up to $m = 9/20$ (top curve). 
}
\end{figure}

\section{Conclusions}

In summary, we have analyzed the spin excitation
spectra 
for two distinct models in a magnetic field: the structurally dimerized chain
and the spin-Peierls chain.
Below a critical field, $h_{c1}$, both systems are 
in a spin liquid phase, composed of interacting
singlet dimers. Above $h_{c1}$,
the spin gap of the structurally dimerized chain closes, whereas 
the spin-Peierls system supports a singlet-triplet gap up to $h_{c2}$.
Therefore, the 
excitation spectra of these two models are quite different in their
incommensurate phases. In the structurally dimerized chain, a
soliton-antisoliton continuum appears at low energies, separated by a 
dimerization
gap $\Delta_{\pm} = 2\delta$ from a second continuum at higher frequencies. 
In the spin-Peierls chain there is a triplet bound state with an onset
at $\Delta_z$, and a higher-energy continuum of states, starting at 
$\Delta_- + \Delta_+ < 2 \Delta_z$. Parts of the triplet band
may thus overlap with the continuum. 
Common features in the spin excitation spectra of these two systems
are (i) an incommensurate, field-dependent modulation
$q = \pi + 2\pi m(h)$ for $h_{c1} < h < h_{c2}$, (ii)
a loss of overall spectral weight with increasing magnetic field, and
(iii) and full spin polarization beyond $h_{c2}$.

Furthermore, there are qualitative changes
in the spectra of the structurally dimerized chain,
depending on the magnitude of the dimerization parameter $\delta$.
The region of large dimerization in the $h - \delta$ phase diagram can be 
understood
within the valence bond picture of weakly interacting dimers,
whereas for small values of $\delta$ the interactions between the dimers are
important, reducing
the ratio of the singlet to the triplet gap and increasing
the bandwidths of the spectral features.

The triplet excitation spectra of the spin-Peierls chain in the incommensurate
phase contain a one-magnon bound state with an onset at $\Delta_z$, and a
soliton-antisoliton continuum with an onset at higher frequencies,
$\omega = \Delta_- + \Delta_+$. The actual dimerization strengths of real
spin-Peierls compounds are typically smaller than the values we have studied
here, which were chosen 
to improve numerical stability. However, the spin
excitation spectra obtained for $H_{sP}$ are qualitatively similar for the 
whole range of $\delta$ we were able to study. 

We wish to thank
A. Honecker, B. Normand, G. Uhrig, and S. Wessel
for useful discussions,
and acknowledge
the Zumberge Foundation for financial support.


\begin{references}
\bibitem{footnote1} We have assumed for convenience that the lattice
modulation is approximately 
sinusoidal. This is in good qualitative agreement with 
calculations which treat the lattice dynamics in a self-consistent
way. However, it should be noted that the actual modulation has a
solitonic form, 
$\delta_r = \delta (-1)^r \tanh((r-r_0-d/2)/\chi) \tanh((r-r_0+d/2)/\chi)$, 
where $r_0$ denotes the position(s) of the largest local magnetization,
$\chi \sim v_s/\Delta_{z}$ is the soliton width, and $d$ is the 
soliton-soliton distance.
We have checked that this form gives results very similar to the 
sinusoidal form, which has been used throughout this paper.
\bibitem{footnote2} The individual gaps of the finite-size clusters 
actually increase as $\delta \rightarrow 0$. However, a finite-size 
scaling analysis shows that in the thermodynamic limit the spin gap 
decreases monotonously as $\delta$ is lowered from 1 to 0.
\bibitem{footnote3} Terms in the Raman operator which commute with the
Hamiltonian contribute to a featureless peak at zero frequency. 
In the limit $\delta = 0$, the full spin Hamiltonian is proportional 
to the Raman operator, leading to a trivial Raman response: 
$I(\omega) = \delta (\omega)$. Here we are interested in the non-trivial
features of $I(\omega)$ away from $\omega = 0$ at finite $\delta$.
\bibitem{footnote4} Typically, a set of poles with decreasing weights
for the higher-energy poles is indicative of an emerging
continuum. This is the case for the three  major poles at higher energies 
in the Raman spectra at zero field, whereas the lowest pole becomes
an isolated bound state as $N \rightarrow \infty$.
\end{references}
\end{document}